\documentclass[twocolumn,aps]{revtex4}
\newcommand{\beq}{\begin{eqnarray}}
\newcommand{\eeq}{\end{eqnarray}}
\begin{document}
\title{\Large \bf Nonzero Mean Squared Momentum of Quarks in the
Non-Perturbative QCD Vacuum$^{*}$ }
\author{Li-Juan Zhou$^{1,2}$, Leonard S. Kisslinger$^{3}$, Wei-xing
Ma$^{4}$}

\vspace{2.0cm}

\affiliation{${1}$ Department of Information and Computing Science,
Guangxi University of Technology, Liuzhou, 545006, P. R. China\\
${2}$ Institute of Particle Physics, Hua-zhong Normal University,
Wuhan, 430079,
 P. R. China \\
${3}$ Department of Physics, Carnegie-Mellon
University, Pittsburgh, PA. 15213, USA\\
${4}$ Institute of High Energy Physics, Chinese Academy of Sciences,
P. O. Box 918 Beijing, 100049, P. R. China }

\date{\today}

\vspace{2.0cm}
\begin{abstract}
The non-local vacuum condensates of QCD describe the distributions
of quarks and gluons in the non-perturbative QCD vacuum. Physically,
this means that vacuum quarks and gluons have nonzero mean-squared
momentum, called virtuality. In this paper we study the quark
virtuality which is given by the ratio of the local quark-gluon mixed
vacuum condensate to the quark local vacuum condensate. The two vacuum
condensates are obtained by solving Dyson-Schwinger Equations of a
fully dressed quark propagator with an effective gluon propagator.
Using our calculated condensates, we obtain the virtuality of
quarks in the QCD vacuum state. Our numerical predictions are consistent
with other theoretical model calculations such as QCD sum rules, Lattice
QCD and instanton models.
\end{abstract}

\maketitle

 \vspace{0.2cm} {\bf PACS number(s)}: 14.65.Bt, 24.85.+p,
12.38.Lg\\

 {\bf Key words}: Quark virtuality, QCD vacuum condensates,
 Non-perturbative QCD.

\vspace{1.0cm}

\section{Introduction}

Quantum mechanics dictates that even "empty" space is not empty, but
rather filled with quantum fluctuations of all possible kinds. In
many contexts, such as in atomic physics, these vacuum fluctuations
are subtle effects which can only be observed by precision
experiments. In other situations, especially when interactions of
sufficient strength are involved, the vacuum fluctuations can be of
substantial magnitude and even "condense" into a non-vanishing
vacuum expectation value of some quantum fields, called vacuum
condensates. These vacuum condensates can act as a medium$^{[1]}$,
which influences the properties of particles propagating through it.

An important example of such a vacuum condensate is the Higgs vacuum
expectation value, which is introduced in the Standard Model of
particle physics to generate the masses of quarks, leptons, and the
gauge bosons ($W^{\pm}$, $Z^{0}$) of the weak interaction. The
vacuum expectation value of the Higgs field, $\langle \phi \rangle =
246 GeV $ , is uniquely determined in the Standard Model. The quark
and lepton masses differ from one another only due to the different
strength of the coupling of each fermion to the Higgs field. At the
same time, the quark masses also receive additional contributions
from the quark and gluon condensates in the QCD vacuum. In fact, the
contribution of the QCD vacuum condensates to the masses for the
three light quarks (u,d,s) considerably exceed the mass believed to
be generated by the Higgs field$^{[2]}$.

The non-vanishing value of chiral quark vacuum condensates signals
the spontaneous breaking of chiral symmetry in QCD, and
quantitatively it is related to the pseudo-Goldstone bosons mass
spectrum$^{[3]}$. Due to non-perturbative effects of QCD, the vacuum
of QCD has a nontrivial structure. The vacuum condensates are very
important in the elucidation of the QCD structure and in description
of hadron properties. If the vacuum acts as a medium and influences
the properties of fundamental particles and their interactions, its
properties can conceivably change. This idea has important
implications in many aspects of physics.

The non-perturbative vacuum of QCD is densely populated by long-wave
fluctuations of quark and gluon fields. The order parameters of this
complicated state are characterized by the vacuum matrix elements of
various singlet combinations of quark and gluon fields, such as
\begin{eqnarray}
&&\langle 0 \mid :\bar{q}q: \mid 0 \rangle,~~~\langle 0 \mid
:\bar{q}[ig_{s}\sigma_ {\mu\nu} G^{a}_ {\mu\nu}
\frac{\lambda^{a}}{2}]q:
\mid 0 \rangle, \nonumber \\
&&~~~\langle 0 \mid
:\bar{q}\gamma_{\mu}\frac{\lambda^{a}}{2}q\bar{q}\gamma_{\mu}\frac{\lambda^{a}}
{2}q:\mid 0 \rangle, \nonumber
\end{eqnarray}
\begin{eqnarray}
\langle 0 \mid :G^{a}_ {\mu\nu}G^{a}_ {\mu\nu}: \mid 0
\rangle,~~~\langle 0 \mid :f^{abc} G^{a}_ {\mu\nu}G^{b}_
{\nu\rho}G^{c}_ {\rho\mu}: \mid 0 \rangle, \cdots,
\end{eqnarray}
which are called vacuum condensates of QCD, where $q(x)$ is the
quark field, $G^{a}_{\mu\nu}$ represents the gluon field strength
tensor with $a$ being color index ($a = 1$,$2$, $\cdots$, $8$), and
can be expressed as
\begin{eqnarray}
G^{a}_ {\mu\nu}(x)= \partial_ {\mu} A^{a}_ {\nu}(x)-\partial_
{\nu}A^{a}_ {\mu}(x) + g_ {s}f^{abc}A^{b}_ {\mu}(x)A^{c}_ {\nu}(x).
\end{eqnarray}
$\lambda^{a}$ in expression (1) is the $SU(3)$ Gell- Mann matrix,
$f^{abc}$ represent the $SU_{c}(3)$ structure constants, and $g_{s}$
in Eq.(2) is the coupling constant related to the so-called QCD
running coupling constant $\alpha_{s}$ by
$\alpha_{s}(Q)=\frac{g^{2}_{s}(Q)}{4\pi}$. $A^{a}_{\mu}$ is the gluon
field, $\sigma_{\mu\nu}=
\frac{i}{2}(\gamma_{\mu}\gamma_{\nu}-\gamma_{\nu}\gamma_{\mu})$ in
Euclidean space with $\gamma_{\mu}$ being a Dirac Matrix.

In QCD by condensates we mean the vacuum mean values $\langle 0 \mid
O_{i} \mid 0 \rangle$ of the local operators $O_{i}(x)$, which arise
due to non-perturbative effects. The latter point is very important
and needs clarification. When determining vacuum condensates one
implies the averaging only over non-perturbative fluctuations. If
for some operators $O_{i}$ the non-zero vacuum mean value appears
also in the perturbative theory, it should not be taken into account
in determination of the condensate. In other words, when determining
condensates the perturbative vacuum mean values should be subtracted
in calculation of the vacuum averages.

Separation of perturbative and non-perturbative contributions to the
quark propagator, $S_q(x)$, has some arbitrariness. For the
nonperturbative propagator, defined in Sect. II, one makes an
expansion of vacuum expectation values involving antiquark-quark
fields that would vanish in a perturbative vacuum, the local vacuum
condensates. The nonzero local quark vacuum condensate $\langle 0
\mid :\bar{q}(0)q(0) : \mid 0 \rangle $ is responsible for the
spontaneous breakdown of chiral symmetry. The nonzero local gluon
vacuum condensate $\langle 0 \mid : G^{a}_{\mu\nu}G^{a}_{\mu\nu}:
\mid 0 \rangle $ defines the mass scale of hadrons through the trace
anomaly$^{[4]}$.

The non-local vacuum condensates $\langle 0 \mid : \bar{q}(x) q(0) :
\mid 0 \rangle $ describe the distribution of quarks in the
non-perturbative vacuum$^{[5]}$. Physically, this means that vacuum
quarks have a nonzero mean-squared momentum called virtuality.
Indeed, the quark average virtuality is connected with the vacuum
expectation values$^{[6,7,8]}$ which will be discussed in the
Sect. III.

 Studying the quark virtuality is of paramount importance for
 present day particle and nuclear physics, since it is not only
 related to the property of QCD vacuum states but also to quark
 vacuum condensates. In this work, we study the quark virtuality in the
 QCD vacuum by calculating quark and gluon vacuum condensates.
 The quark vacuum condensates are obtained by solving
 Dyson-Schwinger Equations (DSEs)$^{[9]}$ of fully dressed quark
 propagators with an effective gluon propagator under the constraints
 of an Operator Product Expansion (OPE)$^{[10]}$. In Sect. II, we briefly
 introduce the DSEs for a fully dressed quark propagator and
 formulae of various quarks vacuum condensates. In Sect. III the  quark
virtuality is defined; and the corresponding formulism of the
virtuality is also derived in this section. In Sect. IV, our
numerical results on the quark virtuality are presented. Our
concluding remarks of this study are given in Sect. V.

\section{DSEs for quark propagator}

To study quark virtuality, we need to know two quark vacuum
condensates, and quark gluon mixed vacuum condensates. Therefore, we
begin with a study of the quark propagators, which determine various
quark condensates and quark gluon mixed vacuum condensates under the
OPE constraints. The quark propagator is defined by
\begin{eqnarray}
S_{q}(x) &=& \langle 0 \mid T[q^a(x)\bar{q}^b(0)] \mid 0 \rangle
\end{eqnarray}
where $q^a(x)$ ($q^{b}(x)$) is a quark field with color $a$ ($b$),
and $T$ is the time-ordering operator. The fully dressed quark
propagator in Eq.(3) can be decomposed into a perturbative part and
a non-perturbative part. In other words, one can write the quark
propagator$^{[11,12]}$ as
\begin{eqnarray}
S_{q}(x) &=& S_{q}^{PT}(x)+S_{q}^{NP}(x),
\end{eqnarray}
where, expanding in the quark mass $m_f$,
\begin{eqnarray}
S_{q}^{PT}(x)&=&(\frac{1}{2\pi^{2}}\frac{\gamma\cdot
x}{x^{4}}-\frac{m_{f}}{2^{2}\pi^{2}x^{2}})\delta^{ab}+\cdots,
\end{eqnarray}
and
\begin{eqnarray}
\label{6}
&& S_{q}^{NP}(x) =-\frac{1}{12} [ \langle 0 \mid : \bar q(x){q}(0): \mid 0
 \rangle ~~~~~\nonumber \\
&& +\gamma_{\mu}\langle 0 \mid : \bar
q(x)\gamma^{\mu}{q}(0): \mid 0 \rangle ]+\cdots \; ,
\end{eqnarray}
in configuration space, with a sum over color. For short distances, the Taylor
expansion of the scalar part of $S_{q}^{NP}(x)$, $\langle 0 \mid :
\bar q(x){q}(0): \mid 0 \rangle$, reads
\begin{eqnarray}
\label{7}
&&\langle 0 \mid : \bar q(x){q}(0): \mid 0 \rangle=\langle 0 \mid :
\bar q(0){q}(0): \mid 0 \rangle \nonumber \\
&&-\frac{x^{2}}{4}\langle 0 \mid : \bar q(0)[ig_{s}\sigma
G(0)]{q}(0): \mid 0 \rangle +\cdots .
\end{eqnarray}
In Eq.(7) the local operators of the expansion are the local quark
vacuum condensates, the quark-gluon mixed condensate, and so forth.

An important observation is that the inverse quark propagator in
momentum space can also be written in Euclidean space as
\begin{eqnarray}
S_{f}^{-1}(p) &=& i \rlap/p \cdot A_{f}(p^{2}) + B_{f}(p^{2}),
\end{eqnarray}
which is renormalized at space - like point $\mu^{2}$ according to
$A_{f}(\mu^{2}) = 1 $ and $ B_{f}(\mu^{2}) = m_{f}(\mu^{2})$, with
$m_{f}(\mu^{2})$ being the current quark mass at re-normalization
point $\mu^{2}$. The subscript $f$ in $A_{f}$ and $B_{f}$ stands for
quark flavor u,d and s.

Except for the current quark mass and perturbative corrections, the
functions $[A_{f}(p^{2})-1]$ and $B_{f}(p^{2})$ are non-perturbative
quantities which we refer to as the vector and scalar propagator
condensates, respectively. The DSEs (in the Feynman gauge) satisfied
by $A_{f}$ and $B_{f}$ then can be written as the set of coupled
equations$^{[6,13]}$.
\begin{eqnarray}
&&[A_{f}(s) - 1 ] s = \frac{1}{3\pi^{3}} g^{2}_{s}
\int_{0}^{\infty}s'd s' \int_{0}^{\pi}\sin ^{2}x D(s,s') \nonumber \\
&&\frac{\sqrt{s s'}A_{f}(s')\cos x}{s'A_{f}^{2}(s') +
B_{f}^{2}(s')}dx,
\end{eqnarray}
\begin{eqnarray}
&&B_{f}(s) = \frac{2}{3\pi^3}g_{s}^{2} \int_{0}^{\infty}s'd
s'\int_{0}^{\pi}\sin ^{2}x D(s,s') \nonumber \\
&&\frac{B_{f}(s')}{s'A^{2}_{f}(s')+ B_{f}^{2}(s')}dx ,
\end{eqnarray}
where $s=p^2$ and $g_{s}^{2}D(s,s')=g_{s}^{2}D(s+s'-2\sqrt{ss'}\cos x)$
is the dressed gluon propagator. Now, our task is to solve this set of
coupled equations, Eqs. (9,10), and get the solutions $A_{f}(s)$ and
$B_{f}(s)$.

One can solve the two coupling integral equations, Eqs. (9,10), using
an effective gluon propagator such as
\begin{eqnarray}
g_{s}^{2}D_{\mu\nu}^{ab}(q) = \delta^{ab} \delta_{\mu\nu} g_{s}^{2}D
(q)=\delta^{ab} \delta_{\mu\nu}\frac{4\pi \alpha (s)}{s},
\end{eqnarray}
where $\alpha(s)$ stands for quark-quark interaction which can be,
for example, well approximated$^{[10]}$ by
\begin{eqnarray}
\alpha(s) = 3\pi s \frac{\chi^{2}}{4 \Delta^{2}}e^{-s/\Delta} +
\frac{\pi d }{\ln( s/\Lambda^{2} + \epsilon)}.
\end{eqnarray}
$\chi$ in Eq. (12) is the strength of the interaction, and $\Delta$
is its range parameter. The first term of Eq.(12) simulates the
infrared enhancement and confinement, and the second term matches to
the leading log renormalization group results. The parameter
$\epsilon$ can be varied in the range $1.0-2.5$. We take
$\epsilon$ to be $2.0$ in the present calculations. The strength
parameter $\chi$ and the parameter $\Delta$ are determined by
fitting the solutions of DSEs to the pion decay constant$^{[12]}$,
and they are listed in table 1.

\begin{center}
{Table 1. Values of the strength parameter $\chi$ and range
parameters $\Delta$ of the quark-quark interaction used in our
present calculations}.

\vspace{0.4cm}
\begin{tabular}{|c|c|c|}\hline
Set no. & Range  $\Delta$ ~~~~~& Strength  $\chi$~~~ \\
\hline
Set 1 ~~~&  0.40 $\rm GeV^{2}$ ~~~  & 1.84 $\rm GeV$~~~ \\
\hline
Set 2 ~~~&  0.20 $\rm GeV^{2}$~~~  & 1.65 $\rm GeV$~~~ \\
\hline
Set 3 ~~~&  0.02 $\rm GeV^{2}$~~~  & 1.50 $\rm GeV$~~~ \\
\hline
\end{tabular}
\end{center}

The QCD scale parameter $\Lambda$ and the value of $d$ with the
flavor number $N_{f}=3$ are given by
\begin{eqnarray}
\Lambda=0.2\rm GeV,~~~d=12/(33-2N_{f})=12/27
\end{eqnarray}

The non-local quark vacuum condensate $\langle 0\mid
:\bar{q}(x)q(0):\mid 0\rangle$ is then given by the scalar part of
Fourier transformed inverse quark propagator$^{[12,13]}$,
\begin{eqnarray}
&&\langle 0\mid :\bar{q}(x)q(0):\mid 0\rangle~~~~~~~~~~~~~~~~~~~~~~~~~~~~~~~~~~~~~~~~~ \nonumber \\
&=&(-4N_{c})\int
\frac{d^{4}p}{(2\pi)^4}\frac{B_{f}(p^{2})e^{ipx}}{p^{2}A^{2}_
{f}(p^{2})+ B^{2}_{f}(p^{2})}~~~~~~~~~~~~ \nonumber\\
&=&-\frac{3}{4\pi^{2}}\int_{0}^{\infty}s ds
\frac{B_{f}(s)}{sA^{2}_{f}(s) + B^{2}_{f}(s)}
\frac{2J_{1}(\sqrt{sx^2})}{\sqrt{sx^2}}
\end{eqnarray}
where the color number $N_{c}=3$. Using the expansion of the $J_1$ Bessel
function, $2 J_1(\sqrt{sx^2})/\sqrt{sx^2}=1-s x^2/8+...$, and the definitions
of the condensates given in Eq(7), one finds that the quark condensate is
\begin{eqnarray}
\label{15} \langle 0\mid : \bar{q}(0)q(0):\mid 0\rangle =
-\frac{3}{4\pi^{2}}\int_{0}^{s_o}ds \frac{s B_{f}(s)}{sA^{2}_
{f}(s) + B^{2}_{f}(s)}
\end{eqnarray}
while the local quark - gluon mixed vacuum condensate, $\langle 0
\mid : \bar{q}(0)[i g_ {s} \sigma G(0)] q(0) : \mid 0 \rangle $ is
\begin{eqnarray}
\label{16}
\langle 0 \mid : \bar{q}(0) [ig_ {s} \sigma G(0)] q(0) : \mid 0 \rangle
 \nonumber \\
=-\frac{3}{8\pi^{2}}\int_{0}^{s_o} ds s^2 \frac{B_
{f}(s)}{sA^{2}_ {f}(s) + B^{2} _{f}(s)}.
\end{eqnarray}

Note that the upper limit of the integrals over s for the DSEs,
Eqs(9,10), is infinity, while for the condensates, Eqs(15,16), there
is a finite limit, $s_o$. That is because the effective gluon
propagator, Eqs(11,12), provides a natural cutoff for the DSEs
integrals, while the range for the condensates is given by the
renormalization point, $\mu$. For light quarks perturbative QCD
begins to dominate nonperturbative QCD at about 3-4 GeV, so a
renormalization point of $\mu^2$ = 10 GeV$^2$ is expected.
Therefore, we use $s_o$ = 10 GeV$^2$. This is explained in detail in
Sect. IV.

  A different derivation of quark - gluon mixed vacuum condensate
explicitly used a form for the nonlocal quark condensate $\langle
0\mid :\bar{q}(x)q(0):\mid 0\rangle = g(x^2)\langle 0\mid
 :\bar{q}(0)q(0):\mid 0\rangle$ (see Ref.$[12]$), and used a model
for $g(x)$ to derive an expression similar to that obtained in the
model used in Ref.$[14]$, but with very different results for the
mixed quark-gluon condensate. In Sect. IV, we use the result for
$g(x)$ from Ref.$[12]$ to estimate errors in our estimate for
$\lambda^2_q$, defined in the following section.

Eqs.(15) and (16) will produce our numerical predictions of local
quark and quark gluon mixed vacuum condensates, which will be then
used to estimate the nonzero mean squared momentum of quarks in the
non-perturbative QCD vacuum.

\section{Nonzero mean squared momentum of quark in QCD vacuum}
The quantities $f(\nu)$ were introduced to represent nonlocal
condensates$^{[15,16]}$. Their explicit form completely determines
the coordinate dependence of the condensates, and describes the
virtuality distribution of quarks in the non-perturbative vacuum.
Its $n$-th moment is proportional to the vacuum expectation value of
the local operator with the covariant derivative squared $D^{2}$ to
the $n$-th power$^{[17]}$:
\begin{eqnarray}
\int_{0}^{\infty}\nu^{n}f_{q}(\nu)d\nu=\frac{1}{\Gamma(n+2)}\frac{\langle
0\mid : \bar{q}(D^{2})^{n}q:\mid 0\rangle}{\langle 0\mid :
\bar{q}q:\mid 0\rangle},
\end{eqnarray}
where the covariant derivative
$D_{\mu}=\partial_{\mu}-ig_{s}A_{\mu}$, with
$A_{\mu}=A_{\mu}^{a}\lambda^{a}/2$ and $\lambda^{a}$ is a $SU(3)$
Gell-mann  matrix. It is natural to suggest that the vacuum
expectation values in the right-hand side of Eq. (17) should exist
for any $n$. The two lowest moments $(n=0, n=1)$ give the
normalization condition $(n=0)$ and the average vacuum virtuality of
quarks $(n=1)$, $\lambda_{q}^{2}$, respectively
\begin{eqnarray}
\int_{0}^{\infty}f_{q}(\nu)d\nu = 1,
\end{eqnarray}
and
\begin{eqnarray}
\label{19}
\int_{0}^{\infty}\nu f_{q}(\nu)d\nu=\frac{1}{2}\frac{\langle 0\mid :
\bar{q}D^{2}q:\mid 0\rangle}{\langle 0\mid : \bar{q}q:\mid
0\rangle}\equiv\frac{\lambda_{q}^{2}}{2}.
\end{eqnarray}
To illustrate the definition of quark virtuality in the
non-perturbative vacuum, $\lambda_{q}^{2}$, let us now consider the
Taylor expansion of the simplest gauge invariant condensate

\begin{eqnarray}
&&\langle 0\mid : \bar{q}(0)E(0,x;A)q(x):\mid 0\rangle
\equiv \langle 0\mid :\bar{q}(0)q(x):\mid 0\rangle \nonumber \\
&=& \sum_{n=0}^{\infty}\frac{1}{n !}x_{\mu_{1}}\cdots
x_{\mu_n}\langle 0\mid : \bar{q}D^{\mu_1}\cdots D^{\mu_n} q:\mid
0\rangle \nonumber \\
&=&\langle 0\mid:\bar{q}q:\mid 0\rangle + \frac{x^2}{8}\langle
0\mid:\bar{q}D^2 q:\mid 0\rangle + \cdots,
\end{eqnarray}
where $E = Pexp [i\int _{x}^{y} A_{\mu}(z)dz^{\mu}]$ is the
path-ordered Schwinger phase factor (the integration is performed
along the straight line ) required for gauge invariance and $
A_{\mu}(z) = A^{a}_{\mu}(z) \lambda ^{a}/2$.

The quantity $\lambda_{q}^{2}$, defined in Eq(\ref{19}),
\begin{eqnarray}
\lambda_{q}^{2}\equiv\frac{\langle 0\mid : \bar{q}D^2 q:\mid
0\rangle}{\langle 0\mid : \bar{q}q:\mid 0\rangle },
\end{eqnarray}
was introduced for an expansion of the nonlocal quark
condensate$^{[5]}$, which can be interpreted as the average
virtuality of the vacuum quarks. Note that the operator $\langle 0
\mid : \bar{q}D^2 q:\mid 0\rangle$ can be presented in a different
form:
\begin{eqnarray}
&~&\langle 0\mid : \bar{q}D^{2} q:\mid 0 \rangle \equiv \langle 0
\mid : \bar{q}D^{\mu}D_{\mu} q:\mid 0 \rangle ~~~~~~~~~ \\&=&
\langle 0\mid
:\bar{q}D^{\mu}g_{\mu\nu} D^{\nu} q:\mid 0 \rangle \nonumber \\
& = & \langle 0 \mid : \bar{q}\rlap/D \rlap/D q:\mid 0 \rangle -
\langle 0 \mid
:\bar{q}D^{\mu}D^{\nu}\sigma_{\mu\nu} q:\mid 0 \rangle \nonumber \\
& = & -m_{q}^{2}\langle 0 \mid : \bar{q}q:\mid 0 \rangle -
\frac{1}{2}\langle 0 \mid :
\bar{q}[D^{\mu},D^{\nu}]\sigma_{\mu\nu}q:\mid 0\rangle \nonumber \\
& = & -m_{q}^{2}\langle 0\mid : \bar{q}q:\mid 0\rangle +
\frac{1}{2}\langle 0\mid :
\bar{q}[ig_{s}G^{\mu\nu}\sigma_{\mu\nu}]q:\mid 0\rangle, \nonumber
\end{eqnarray}
where $\rlap/D=\gamma^{\mu}D_{\mu}$, and we have used the identity
\begin{eqnarray}
g_{\mu\nu}=\gamma_{\mu}\gamma_{\nu}-\frac{\gamma_{\mu}\gamma_{\nu}-
\gamma_{\nu}\gamma_{\mu}}{2}=\gamma_{\mu}\gamma_{\nu}-\sigma_{\mu\nu},
\end{eqnarray}
the equation of motion
\begin{eqnarray}
\rlap/Dq(x)=-im_{q}q(x),
\end{eqnarray}
and the definition of the field strength tensor
\begin{eqnarray}
[D_{\mu},D_{\nu}]=-igG_{\mu\nu}.
\end{eqnarray}
Thus, the "average virtuality" of the vacuum quarks $\langle 0\mid :
\bar{q}D^2 q:\mid 0\rangle$ is directly related to the "average
vacuum gluon field strength" $\langle 0 \mid : \bar{q}[ i g_{s}
G^{\mu\nu}\sigma_{\mu\nu}] q : \mid 0 \rangle $. In many
papers$^{[18]}$ one can find the notation
\begin{eqnarray}
\langle 0\mid : \bar{q}[ig_{s}G^{\mu\nu}\sigma_{\mu\nu}]q:\mid
0\rangle \equiv m_{0}^{2}\langle 0\mid : \bar{q}q:\mid 0\rangle.
\end{eqnarray}
Using Eqs. (21,22,26), one can write the vacuum quark virtuality
$\lambda_{q}^{2}$ as
\begin{eqnarray}
\lambda_{q}^{2}=\frac{m_{0}^{2}}{2}-m_{q}^{2}.
\end{eqnarray}
Namely,
\begin{eqnarray}
\label{29} \lambda_{q}^{2}=\frac{1}{2} \frac{\langle 0\mid :
\bar{q}[ig_{s}G^{\mu\nu}\sigma_{\mu\nu}]q:\mid 0\rangle}{\langle
0\mid : \bar{q}q:\mid 0\rangle }-m_{q}^{2}.
\end{eqnarray}
For light quarks, the mass $m_{q}^{2}$ term is so small that it can
be neglected. Finally, we arrive at
\begin{eqnarray}
\lambda_{q}^{2} \simeq \frac{1}{2} \frac{\langle 0\mid :
\bar{q}[ig_{s}G^{\mu\nu}\sigma_{\mu\nu}]q:\mid 0\rangle}{\langle
0\mid : \bar{q}q:\mid 0\rangle }.
\end{eqnarray}
Eq. (29) is starting point of our calculations on quark virtuality.
As we see from Eq. (29), in order to get $\lambda_{q}^{2}$ we have
to calculate quark-gluon mixed vacuum condensate $\langle 0\mid :
\bar{q}[ig_{s}G^{\mu\nu}\sigma_{\mu\nu}]q:\mid 0\rangle$ and two
quark vacuum condensate $\langle 0\mid : \bar{q}q:\mid 0\rangle$ by
the use of Eqs. (15) and (16).

\section{Numerical results and error estimate}
Using the solutions of DSEs of Eqs. (9,10), $A_{f}$ and $B_{f}$,
with three different sets of the quark-quark interaction parameters
given in Table 1, leads to our following theoretical predictions for
the local two quark vacuum condensates, local quark-gluon mixed
vacuum condensates via Eqs. (15,16). The corresponding theoretical
results are listed in Table 2-3.
\begin{center}
{Table 2. The local two quark vacuum condensates of QCD, $\langle 0
\mid :\bar{q}q:\mid 0 \rangle_{\mu}^{f}$, $f$ stands for quark
flavor and $\mu$ denotes renormalization point, $\mu^2$=10 GeV$^2$}.

\vspace{0.4cm}
\begin{tabular}{|c|c|c|c|}\hline
 Set No. & $\langle 0\mid :\bar{q}q:\mid 0\rangle^{u,d}
$ & $\langle 0\mid :\bar{q}q:\mid
0\rangle^{s}$ \\
\hline Set 1 & $-0.013(\rm GeV)^{3} $ & $-0.071(\rm GeV)^3$  \\
\hline Set 2 & $-0.0078 (\rm GeV)^{3}$ & $-0.068(\rm GeV)^3 $  \\
\hline Set 3 & $-0.0027 (\rm GeV)^{3}$ & $-0.065(\rm GeV)^3 $  \\
\hline
\end{tabular}
\end{center}

\begin{center}
{ Table 3. The local quark-gluon mixed vacuum condensates, $\langle
0\mid :\bar{q} [i g_{s}\sigma G] q :\mid 0\rangle_{\mu}^{f}$},
$\mu^2$=10 GeV$^2$.

\vspace{0.4cm}
\begin{tabular}{|c|c|c|}\hline
Set No. & $\langle 0\mid :\bar{q} [i g_{s}\sigma G] q :\mid 0\rangle
^{u,d}$ & $\langle 0\mid :\bar{q}
[ig_{s}\sigma G]q :\mid 0\rangle ^{s}$ \\
\hline Set 1 & $-0.015 (\rm GeV)^{5}$ & $-0.186 (\rm GeV)^5$ \\
\hline Set 2 & $-0.010 (\rm GeV)^{5}$ & $-0.189 (\rm GeV)^5$ \\
\hline Set 3 & $-0.0078 (\rm GeV)^{5}$ & $-0.193 (\rm GeV)^{5}$ \\
\hline
\end{tabular}
\end{center}

Our results for two quark local vacuum condensates are consistent
with the predictions by Gall-Mann-Oakes-Renner relation
(GMOR)$^{[19,20]}$, $(m_{u}+ m_{d})\langle 0\mid :\bar{q}q :\mid
0\rangle=-\frac{1}{2}m_{\pi}^{2}f_{\pi}^{2}$, where $m_{u}$ and
$m_{d}$ are current quark masses with value of $m_{u}+m_{d}=9.7 \rm
MeV^{[20]}$, and $m_{\pi}=140 \rm MeV$, $f_{\pi}=93 \rm MeV$ are the
mass and decay constant of a pion, respectively. Substituting these
values into the GMOR relation produces $\langle 0\mid :\bar{q}q :
\mid 0\rangle=-0.0087 GeV^{3}$, which is reasonably consistent with
the u,d quark condensates shown in Table 2, within errors.

Our theoretical results in Table 2 and 3 are also consistent with
the empirical values used widely in QCD sum rules$^{[21]}$, with
$\langle 0\mid :\bar{q}q :\mid 0\rangle^{u,d} \simeq -0.013$
GeV$^3$, as we obtained with Set 1. We also obtained a result 
consistent with the predictions of Lattice calculations$^{[22]}$.

Using Eq.(29) and our numerical predictions of the two quark local
vacuum condensates, and the local quark-gluon mixed vacuum
condensates given respectively in tables 2 and 3, the quark
virtuality,  the nonzero mean squared momentum of quarks in
non-perturbative QCD vacuum state, is given by
\begin{eqnarray}
\lambda^{2}_ {u,d}&=& \frac{1}{2} \frac{\langle 0 \mid : \bar{q}(0)
[i g_ {s}\sigma_ {\mu\nu} G^{a}_ {\mu\nu}\frac{\lambda^{a}}{2}]q(0)
: \mid 0 \rangle_  {u,d}}{\langle 0\mid : \bar{q}(0)q(0):\mid
0\rangle_ {u,d}} \nonumber \\
&&= 0.57 \rm GeV^{2},
\end{eqnarray}
for $u$, $d$ quark, Set 1.

For $s$ quark, we obtained
\begin{eqnarray}
&&\lambda^{2}_ {s} = \frac{1}{2} \frac{\langle 0 \mid : \bar{q}(0) [i
g_{s}\sigma_ {\mu\nu} G^{a}_ {\mu\nu}\frac{\lambda^{a}}{2}]q(0) :
\mid 0 \rangle_ {s}}{\langle 0 \mid : \bar{q}(0)q(0):\mid 0\rangle_
{s}} \nonumber \\
&& = 1.31 \rm GeV^{2},
\end{eqnarray}
for Set 1.

In the summary of this section, our predictions of $\lambda^{2}_
{u,d}$ and $\lambda^{2}_ {s}$ with three different sets of
quark-quark interaction parameters given in table 1 are listed in
Table 4.

\begin{center}
{ Table 4. The virtualities of three light quarks $u, d$ and $s$ in
the QCD vacuum state}.

\vspace{0.4cm}
\begin{tabular}{|c|c|c|}\hline
Set No. & $\lambda^{2}_ {u,d} \rm [GeV]^{2}$ & $\lambda^{2}_
{s}\rm [GeV]^{2}$ \\
\hline Set 1 & $0.57$ & $1.31$ \\
\hline Set 2 & $0.68$ & $1.38$ \\
\hline Set 3 & $1.43$ & $1.48$ \\
\hline
\end{tabular}
\end{center}

All our theoretical results are in an acceptable range$^{[23]}$ of
$\lambda^{2}_{q}$ between $0.4 \sim 2.50 GeV^{2}$. For example, for
$u$ and $d$ quarks the standard QCD sum rule estimation$^{[24]}$
gives $\lambda^{q}_{u,d}= 0.4\pm 0.1 GeV^{2}$, the QCD sum rule
analysis of pion form factor$^{[25]}$ produces $\lambda^{q}_{u,d} =
0.70 GeV^{2}$, and Lattice QCD calculations$^{[26]}$ predicts
$\lambda^{2}_{u,d}=0.55 GeV^{2}$. For $s$ quark, Lattice QCD
$^{[26]}$ gives $\lambda^{2}_{s}=2.50 GeV^{2}$, and the instanton
model prediction$^{[27]}$, $\lambda^{2}_{s}=1.40 GeV^{2}$. The
$\lambda^{2}_{u,d} =1.43$ for set 3 is larger than that for set 1
($0.57$) and set 2 ($0.68$). The reason is that the range parameter
$\Delta$ is an order of magnitude smaller. Therefore we observe that
all our predictions are in a good agreement with the calculations
cited by Refs. [23-27], but our method of calculation is quite different
from others. However, it should be also pointed out that both the
condensates and virtualities depend on renormalizatin point
$\mu^{2}$. We discuss this dependence and estimated error of
$\lambda^{2}_{q}$ in the following.

\subsection{Dependence of $\lambda^{2}_{q}$ on $\mu^{2}$ }

As it has been mentioned in Sect. I, in QCD by condensates we mean
the vacuum mean values $\langle 0\mid O_{i}\mid 0\rangle$ of the
local operators $O_{i}$, which arise due to non-perturbative
effects. When determining vacuum condensates one implies the
averaging only over non-perturbative fluctuations. If for some
operators $O_{i}$ the non-zero vacuum mean value appears also in
perturbation theory, it should not be taken into account in the
determination of the condensates. In other words, when determining
condensates the perturbation vacuum mean values should be subtracted
in calculation of the vacuum averages.

Separation of perturbation and non-perturbation contribution into
vacuum mean values has some arbitrariness. Usually, this
arbitrariness is avoided by introducing some renormalization point
$\mu^{2}$. Integration over momenta of virtual quarks and gluons in
the region below $\mu^{2}$ is referred to condensates, above
$\mu^{2}$ is referred to perturbation theory. In such a formulation
condensates depend on the renormalization point $\mu^{2}$, $\langle
0 \mid O_{i} \mid 0\rangle_{\mu^{2}}$. Therefore, $\lambda^{2}_{q}$
depends on the renormalization point $\mu^{2}$, which we choose as
10 GeV$^2$ as explained in the paragraph following Eqs(15,16). This
is consistent with our conclusions regarding our current calculations.

\subsection{Estimate of Errors in $\lambda^{2}_{q}$}

Our values of $\lambda^{2}_q$ for the u,d and s quarks are given by
the quark condensate, $\langle 0\mid :\bar{q}(0)q(0):\mid 0\rangle$,
and the quark-gluon mixed vacuum condensate, $\langle 0\mid
:\bar{q}[ig_{s}G^{\mu\nu}\sigma_{\mu\nu}]q:\mid 0\rangle$. The quark
condensate is known to about 10 \%, but as discussed in Sect. II,
approximations and models are needed to estimate the quark-gluon
mixed condensate, and there can be some errors, depending on the
model used.

Although it is difficult to give an accurate estimate of the errors,
from the results for the condensates and virtualities given in
Tables 2, 3, and 4 one sees that the virtuality is estimated to
about a factor of two. Since the parameters for Sets 1, and 2 given
in Table 1 are more reasonable than Set 3, one may estimate from
Table 4 that our final results for the virtuality are within about
20 \%.

\section{Concluding remarks}
We study the quark virtuality in the QCD vacuum based on the fully
dressed confined quark propagator described by DSEs. The quark
virtuality is determined by the ratio of the local quark-gluon mixed
vacuum condensates $\langle 0 \mid : \bar{q} [i g_{s} G_
{\mu\nu}\sigma_ {\mu\nu}]q : \mid 0 \rangle$ to local quark vacuum
condensates $\langle 0 \mid : \bar{q}q:\mid 0\rangle$. The local
quark vacuum condensate and local quark gluon mixed vacuum
condensate are obtained by solving the Dyson-Schwinger Equations in
the "rainbow" approximation with an effective gluon propagator in
Euclidean space and the Feynman gauge. The effective gluon
propagator consists of two terms with two parameters: the strength
of interaction $\chi$ and its range $\Delta$. The first term of the
gluon propagator simulates the infrared enhancement and confinement,
and the second term matches to the leading log renormalization group
results. Our calculated results of local quark vacuum condensates
and quark gluon mixed vacuum condensate are in good agreement with
other theoretical model predictions such as QCD sum
rules$^{[24,25]}$, Lattice QCD$^{[26]}$ and instanton
model$^{[27]}$.

Using the numerical results of our present calculations of local
quark and quark-gluon mixed vacuum condensates, the virtualities
$\lambda_{q}$ for light quarks ( $u$, $d$ and $s$ ) are obtained for
three different sets of parameters $\chi$ and $\Delta$. The results
are given in table 4. We find numerically that the contribution from
the second term of gluon effective propagator in
$g^{2}_{s}D^{ab}_{\mu\nu}$, Eq. (12), can be neglected. The dominant
contribution to quark virtuality is from the first term of Eq. (12).

In conclusion, we predict the quark virtuality using a different
method: solving DSEs, and using its numerical solutions $A_{f}$ and
$B_{f}$. Our theoretical results are consistent with all
calculations of QCD sum rules, Lattice QCD and instanton models.
However, it should be noticed that the predictions depend on
renormalization point $\mu^{2}$, the separation between perturbative
and non-perturbative part of QCD, since the vacuum condensate
average only over non-perturbative vacuum fluctuations, and the
perturbative contribution must be subtracted from any calculations.
The detailed discussion on $\mu^{2}$ dependence will be published in
our forthcoming paper. We believe this study is very important for
investigation of QCD vacuum properties, and has many important
applications both in particle physics and in nuclear physics.

\vspace{0.4cm}

{\bf $^{*}$Acknowledgments:} The work was supported in part by
National Natural Science Foundation of China under project No.
10647002; Guangxi Science Foundation for young researchers: 0991009;
Department of Guangxi Education: 200807MS112, and by the NSF/INT
grant number 0529828, USA.


\vspace{1.0cm}

\begin{thebibliography}{99}
\bibitem{1} T. D. Lee, N. Y. Trans, Acad. Sci. {\bf 40} (1980)111.
\bibitem{2} P. W. Higgs, Phys. Rev. Lett. , {\bf 12} (1964) 132; P. W.
Higgs, Phys. Rev. {\bf 145} (1966) 1156; Berndt Muller, "From
quark-gluon plasma to the perfect liquid ", arXiv: nucl-th/0710.3366;
Acta. Phys. Polon. {\bf B 38} (2007) 3705.
\bibitem{3} T. Bechnke et al, Prog. Part. Nucl. Phys., {\bf 48}
(2002) 363; Bing An-li, arXiv: hep-ph/9808441.
\bibitem{4} A. W. Thomas and W. Weise, "Scale invariance and the trace
anomaly",
The structure of the nucleon, pp.170-172, John wiley and Sons,
Singapore, 2000.
\bibitem{5} S.V. Mikhailov and A.V. Radyushkin, JETP Lett. {\bf 43} (1986)
712.
\bibitem{6} Zhou Li-juan, Ma Wei-xing, Chinese Physics Letters, {\bf 21} (2004)
1471; Zhou Li-juan and Ma Wei-xing, Chinese Physics Letters, {\bf 20}
(2003) 2137; A. E. Dorokhov, S. V. Radyushkin, Physics of Partical
and Nuclei(suppl.) {\bf 32} (2001) 554.
\bibitem{7} A. E. Dorokhov, S. V. Esaibegyan,
S. V. Mikhailov, Phys. Rev., {\bf D 56} (1997) 4062.
\bibitem{8} Zhou Li -juan and Ma Wei-xing, Commun. Theor. Phys.,
 {\bf 45} (2006) 1085.
\bibitem{9} F. J. Dyson, Phys. Rev., {\bf75} (1949)1736; L. S. Schwinger,
Proc. Nar. Acad. Sci., {\bf 37} (1951) 452; C. D. Roberts, prog.
Part. Nucl. Phys., {\bf 45} (2000) 511; Zhou Li-juan, Ping Rong-gang,
Ma Wei-xing, Commun. Theor. Phys., {\bf 40} (2003) 558.
\bibitem{10} K. G. Wilson, "On products of quantum fields operators
at short distances" ( Cornell Report, 1964), cornell. (sec. 4.1); R.
A. Brandt, Ann. Phys., {\bf 44} (1967) 221; W. Hubschmid, S. Mallik,
Nucl. Phys., {\bf B 207} (1982) 29.
\bibitem{11} Zhou Li-juan, Ping Rong-gang, Ma Wei-xing, Commun. Theor. Phys.,
{\bf 42} (2004) 875.
\bibitem {12} L. S. Kisslinger, T. Meissner,
Phys. Rev., {\bf C 57} (1998) 1528; M. R. Frank and T. Meissner,
Phys. Rev., {\bf C 53} (1996) 2410; arXiv: hep-ph/9511016; L. S.
Kisslinger, M. Aw, A. Harey, and O. Linsuain, Phys. Rev. {\bf C 60},
(1999) 065204.
\bibitem {13} He Xiao-rong, Zhou Li-juan, Ma Wei-xing, Commun.
Theor. Phys., {\bf 45} (2005)  670.
\bibitem{14} T. Meissner, Phys. Lett. {\bf B 405} (1997) 8.
\bibitem{15} Ma Wei-xing, Zhou Li-juan, Ping Rong-gang, Commun. Theor. Phys.,
{\bf 44} (2005) 333; L. S. Kisslinger, "2004 Review of light
cone field theory", Int. J. Mod. Phys. {\bf E 13} (2004) 375.
\bibitem{16} A. P. Bakulev and A. V. Radyushkin, Phys. Letts. {\bf B271}
(1991) 223; S. V. Mikhailov and A. V. Radyushkin, Sov. J. Nucl.
Phys., {\bf 49} (1989) 493.
\bibitem{17} S. V. Mikhailov and A. V. Radyushkin, Phys. Rev. {bf D 45}
(1992) 1754.
\bibitem{18} V. M. Belyaev and B. L. Ioffe, Zh. Eksp. Teor. Fiz.,
{\bf 83} (1982) 876 [Sov. Phys. JETP, {\bf 56} (1982) 493]; A. A.
Ovchinnikov and A. A. Pivovarov, Yad. Fiz. {\bf 48} (1988) 1135
[Phys. at. Nucl. {\bf 48} (1988) 721].
\bibitem{19} S. N. Nikolaev and A. V. Radyushkin, Nucl. Phys. {\bf B213}
(1983) 285; A. G. Grozin, Intern. J. Mod. Phys. {\bf A 10} (1995) 3497.
\bibitem{20} M. Gell-Mann, R. J. Oakes and B. Renner, Phys. Rev., vol. 175
(1968)
2195; J. C. Collins, A. Dunean and S. D. Joglekar, Phys. Rev. {\bf D 16}
(1977) 438; R. Tarrach, Nucl. Phys., {\bf B 196} (1982) 45.
\bibitem{21} H. G. Dosch and S. Narison, Phys. Letts, {\bf B 417} (1998)
173; M. Kremer and G. Schierholz, Phys. Letts., B 194 (1987)283;
D. B. Leinweber, "QCD sum Rules for Skeptics", DOE/Er/40427-17-N95.
\bibitem{22} L. Giusti et al., Nucl. Phys.,{\bf B 538} (1999) 249; Phys.
Rev., {\bf D 64} (2001) 114508; P. Hernandez, et al., Nucl. Phys.
Proc. (suppl.) {\bf 106} (2002) 766; M. Jamin, J. A. Older and A.
Pich, Eur. Phys. J. {\bf C 24} (2002) 237; L. M.Jamin, HD-THEP-0201,
arXiv: hep-ph/0201174; M. V. Polykov and C. Weiss, Phys. Letts. {\bf
B 387} (1996) 841.
\bibitem{23}V. A. noikov, M. A. Shifman, V. I. Vainshtenl, et al.,
Nucl. Phys. {\bf B237} (1984) 525; L. S. Kisslinger, arXiv:
hep-ph/9906457.
\bibitem{24} C. D. Roberts and G. A. Williams, Prog. Part. Nucl. Phys.
{\bf 33} (1994) 477; T. Meissner, Phys. Lett. {\bf B 405} (1997) 8;
M. A. Bentez, R. M. Barnett, and C. Caso, Particle Data
Group, Particle properties data booklet, April (1990), Phys. Letts.
{\bf B 239} (1990) 412; V. M. Belyaev and B. L. Ioffe, Sov. Phys.
JEPT {\bf 56} (1982) 63c, P. C. Tandy, Prog. Part. Nucl. Phys. {\bf
39} (1997) 117.
\bibitem{25} L. S. Kisslinger and O. Linsuain, arXiv: hep-ph /0110111;
L. S. Kisslinger and M. A. Harly, arXiv: hep-ph/9906457; V. A.
Novikov, M. A. Shifman, V. I. Vainshtein, M. B. Voloshin and V. I.
Zakharov, Nucl. Phys. {\bf B 237} (1984) 525.
\bibitem{26} D. Takumi, I. Noriyoshi, O. Makoto and Hido Suganuma,
Nucl. Phys. {\bf A 721} (2003) 934c; V. M. Belyaev and B. L. Ioffe,
Sov. Phys. JEPT {\bf 56} (1982) 493; C. D. Roberts, R.T. Cahill, M.
E. Sevior and N. Iannelle, Phys. Rev. {\bf D 49} (1994) 125.
\bibitem{27} M. V. Polykov and C Weiss, Phys. Lett. {\bf B387}
(1996) 841.

\end{thebibliography}
\end{document}